\documentstyle[epsf,11pt,amssymb]{article}
\newcommand{\bm}[1]{\mbox{\boldmath$#1\!$}}
\begin{document}
\baselineskip=5mm


\vspace*{0.2cm}

\begin{center}
\noindent {\huge \bf  Plasma emission and radiation  from ultra- \\*[0.1cm]
relativistic Brunel electrons in femtosecond  \\*[0.40cm] laser-plasma  interactions} 

\vspace*{8mm}

\noindent \hspace*{1.9cm} {\LARGE  {\bf R Ondarza-Rovira}$^{1}$  and {\bf TJM Boyd}$^{2}$ }

\

\vspace*{0.3mm}

\noindent \hspace*{0.0cm}  $^{1}${Instituto Nacional de Investigaciones Nucleares, 
A.P.\ 18-1027, M\'{e}xico 11801, DF, Mexico} 

\vspace*{0.1cm}

\noindent \hspace*{0.0cm} $^{2}${Centre for Theoretical Physics, University of 
Essex, Wivenhoe~Park, Colchester ~CO4~3SQ, UK}
\end{center}

\vspace*{0.2cm}


\begin{center}
\parbox{14cm}{\small A highly intense femtosecond laser pulse incident on a plasma target of
supercritical density, gives rise to reflected high-order harmonics of the laser frequency. The radiation model adopted
here considers Brunel electrons -those reinjected into the plasma after performing  a vacuum 
excursion- perturbed by a localized turbulent region of an  electrostatic field that is generated during the interaction,
and characterized by a soliton-like structure. The observed power
spectrum is characterized by a power-law decay scaled as $P_m \sim m^{-p}$, where $m$ denotes the harmonic order. 
In this work an appeal is made to a radiation mechanism from a single particle model that shows harmonic power decays
described -as previously reported from particle-in-cell simulations- within the range $2/3 \leq p \leq 5/3$. Plasma
emission is strongest for values of the similarity parameter $S=(n_e/n_c)/a_0$ in the range $1 \leq S \leq 5$, and
where $n_e$  and $n_c$ are the ambient electron plasma density and the critical density, respectively, and $a_0$ is the
 normalized  parameter of the electric field of the incident light. It was found that the radiation spectra obtained from 
the single particle model here presented is consistent with previously reported power-law 5/3 decays from 
particle-in-cell simulations.}  
\end{center}

\vspace*{5mm}

\noindent {\large \bf 1.\ Introduction}

\

\noindent  High harmonic generation in relativistic interactions from femtosecond laser pulses on solid
density targets has stimulated interest in a number of potential applications in plasma diagnostics and 
technological developments in ultrafast optics for studying dynamical processes in matter, and more recently as 
sources for generating ultrashort-intense pulses in the attosecond range [1].
 
It has long been known from experiments and particle simulations that laser-generated harmonics from the interaction 
of a highly intense light pulse with overcritical density plasmas are characterized by a power-law decay that 
can extend over hundreds or even thousands of harmonic orders. The spectral decay has been 
characterized in different approximations from the laser-plasma parameters [2,3]. In the mildly relativistic regime 
Gibbon [2] first proposed an empirical relation from PIC simulations  for the harmonic efficiency $\eta$, and showed
that this scaled as $\displaystyle \eta \sim 17.2\, a_0^4\, m^{-p}$, where the decay index $p=5$, and $a_0=  8.5\, 
{ \left( I_{20}\, \lambda_{L}^{2} \right) }^{1/2}$  is the normalized quiver momentum, with $I_{20}$ denoting the
 light intensity in units of $10^{20}$ Wcm${}^{-2}$ and $\lambda_L$ the laser wavelength measured in microns. As
values of $a_0$ increased above $a_0=1$, relativistic effects arise in
the interaction physics. In picosecond-pulse experiments, the harmonic spectra was observed 
to span a range of values of $p$ for increasing intensities, from $p=5.5$ at $a_0 \sim 0.64$ to 
$p=3.38$ at $a_0 \sim 2.85$ [3]. More recently, Baeva {\it et al.\ }[4] proposed a model based on a 
similarity model, that the underlying physics is similar when the similarity parameter $S$ remains
constant, with an associated so-called ``universal" value 8/3  to the harmonic power decay. Furthermore, the 
$S$-similarity  analysis predicts a cut-off in the harmonic spectra at $\displaystyle m= \sqrt{8\, \alpha}\, 
\gamma_s^3$, where $\gamma_s$ is the relativistic factor of the plasma surface and $\alpha$ is of order one. No
clear-cut evidence has been found for the spectral cut-off experimentally, though the spectral power decay was
shown to be fitted by indexes  in the range $2.3 \lesssim p \lesssim 2.7$, with the upper limit corresponding to the
predicted ``universal" value.

\

\noindent {\large \bf 2.\ Plasma effects}

\

\noindent In mildly relativistic interactions, Boyd and Ondarza-Rovira (BOR) found evidence of 
emission at the plasma line and at its harmonics [5]. In the UR regime, the ponderomotive force 
of the Langmuir field is strong enough to govern plasma behaviour over time scales $t > \omega_p^{-1}$
and highly localized electrostatic fields are generated in correspondence with density structures
inside the plasma. These structures were identified as sources of emission at the plasma frequency 
and at  its harmonics. Through PIC simulations BOR explored the harmonic power decay in UR  
laser-plasma interactions [6] and spectral decays characterized by an index $p=5/3$ were found and, at higher
 intensities, by $p=4/3$ for $p$-polarized light. When the incident laser pulse is normal 
to the plasma surface, the spectrum is invariably characterized by an 8/3 power law decay, regardless of the
combination $(a_0, n_e/n_c)$. Contrasting this, plasma effects are evident 
in the power harmonic decay for oblique incidence, since the excitation of plasma oscillations
enable the enhancement of emission with lower harmonic power decay indexes, resulting in  
much flatter spectra. As an example, the plasma oscillations for the interaction $a_0=10, n_e/n_c=40$ and
a pulse length $\tau_p=17$ fs are shown in Fig.\ 1(a) around the strong electrostatic field generated, as obtained from 
a PIC  simulation. Figure 1(b) shows the corresponding density contours where the high plasma density spot 
represents a region of strong electrostatic field generation. Also Brunel trajectories of electrons can be discerned
travelling deep inside the plasma.

\renewcommand{\thefigure}{1}
\begin{figure}[h]
\centering \leavevmode
\epsfxsize=8.0cm \epsfysize=5.0cm \epsffile{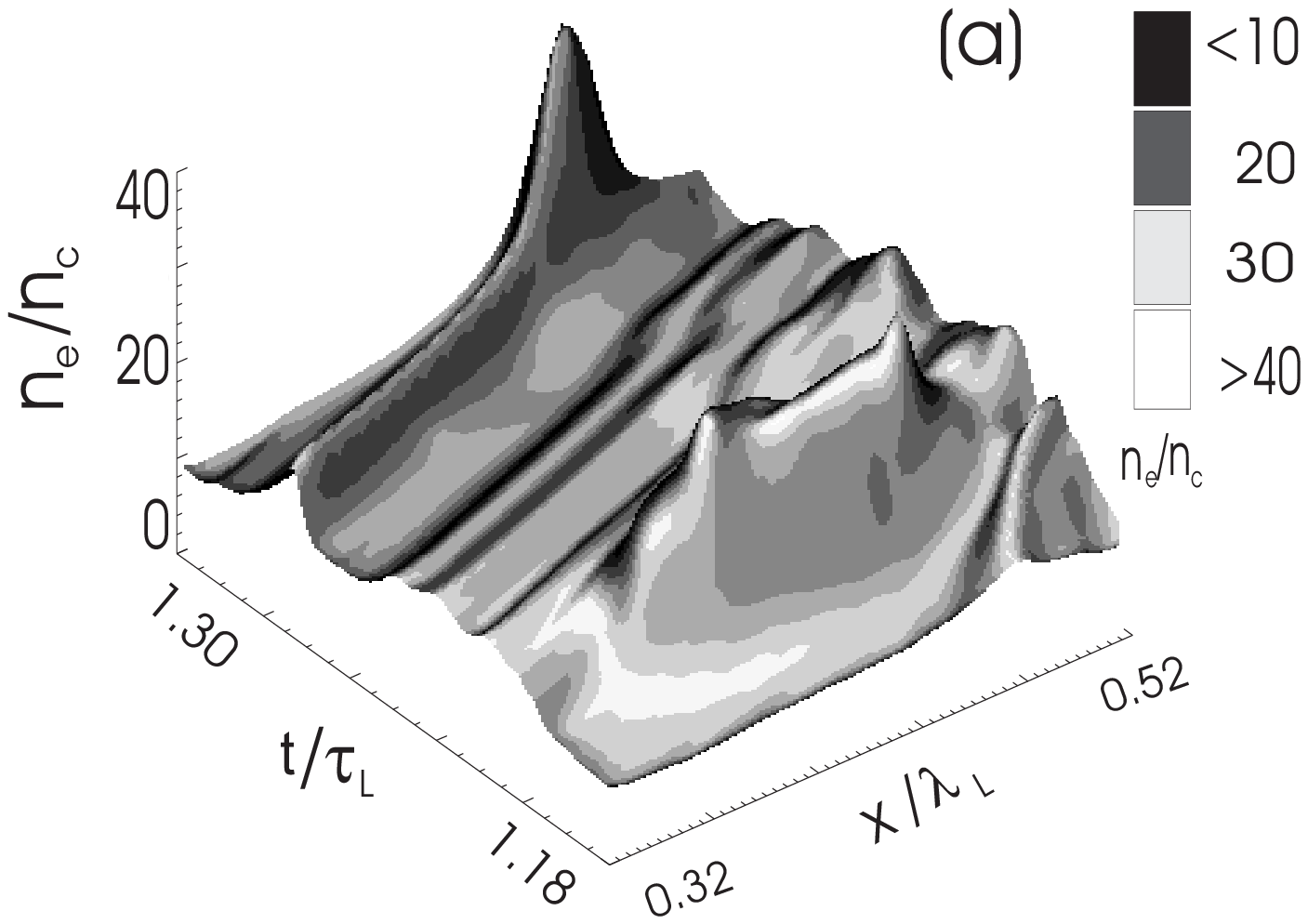} \hspace*{1.0cm}  
\epsfxsize=7.0cm \epsfysize=5.0cm \epsffile{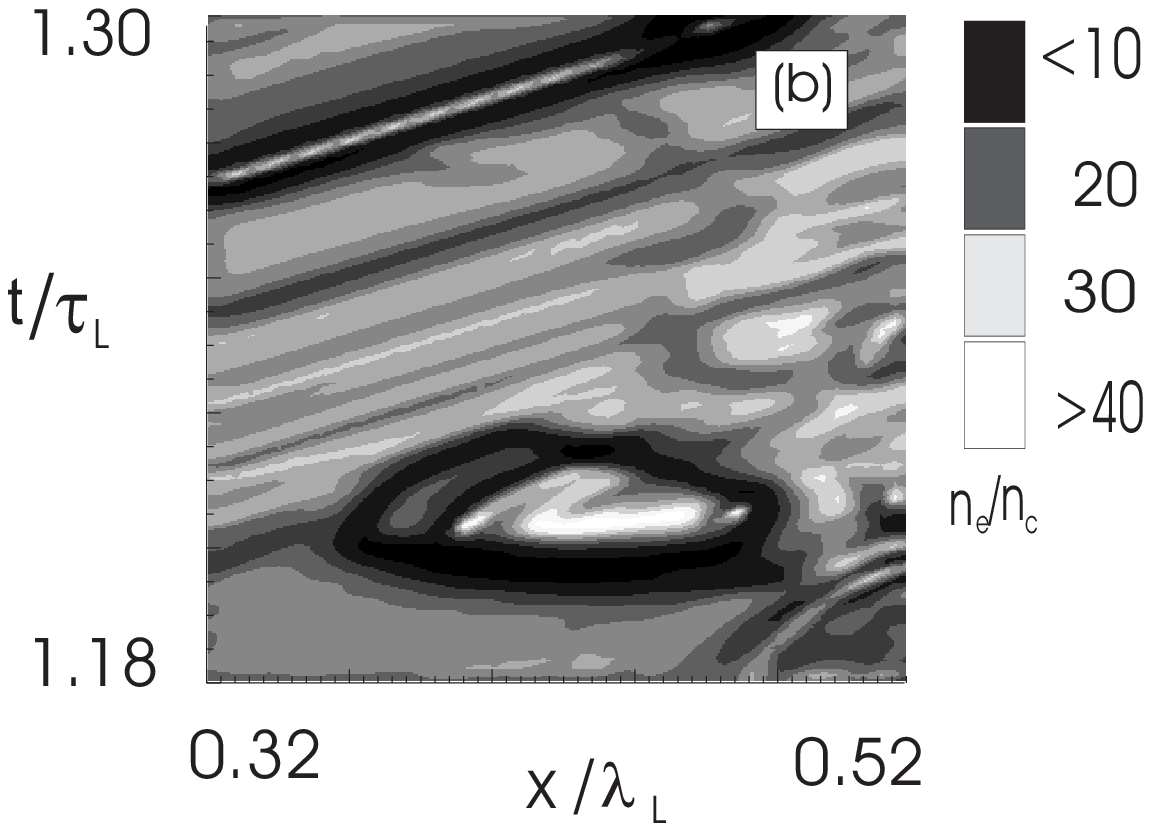} \vspace*{0.1cm} 
\parbox{16.0cm}{\vspace*{0.3cm}Figure 1: Regions in the plasma where the emission source is correlated: 
(a) plasma electron density oscillations and (b) electron density contours at sites of high electrostatic field generation
 inside the plasma.}
\end{figure}

The Brunel electron trajectories were also observed to be correlated with the electrostatic fields in places of beam
 crossings, as shown in Fig.\ 2(a). For $p$-polarization field amplification is also observed (Fig.\ 2(b)), being an order 
of magnitude greater than the $s-$polarized case. These reflected attosecond spikes correlate with the electron
 bunching where strong electrostatic fields are found, and identified as sources of emission [6]. 

 \

\noindent {\large \bf  3.\ Plasma and XUV harmonic emission from ultra-relativistically laser-accelerated electrons 
perturbed by soliton-like electrostatic plasma fields}

\

\noindent In the following we make use of a simple dynamical models to explore the plasma effects on
 harmonic radiation emission from a charged particle relativistically accelerated by a strong electromagnetic pulse, and
 perturbed  by a soliton-like electrostatic field generated in the plasma during the laser-plasma
  interaction -as discussed in the previous section. 

\noindent  Radiation from electron bunching in large-amplitude Langmuir waves has been considered previously in
cases where these have been excited in beam-plasma interactions. In particular, Weatherall and Hobbs [7] showed 
that electrostatic bunching was an effective means of generating plasma harmonics. Subsequently, Weatherall and 
Benford (WB) [8] modelled these fields as soliton-like structures generated inside the plasma and showed that
these sources gave rise to strong nonlinear perturbations when high energy electron beams traverse such structures
resulting in broadband bremsstrahlung emission. This model describes the radiation  effects pro-

\newpage

\vspace*{0.1cm}

\renewcommand{\thefigure}{2}
\begin{figure}[h]
\centering \leavevmode
\epsfxsize=6.0cm \epsfysize=5.0cm \epsffile{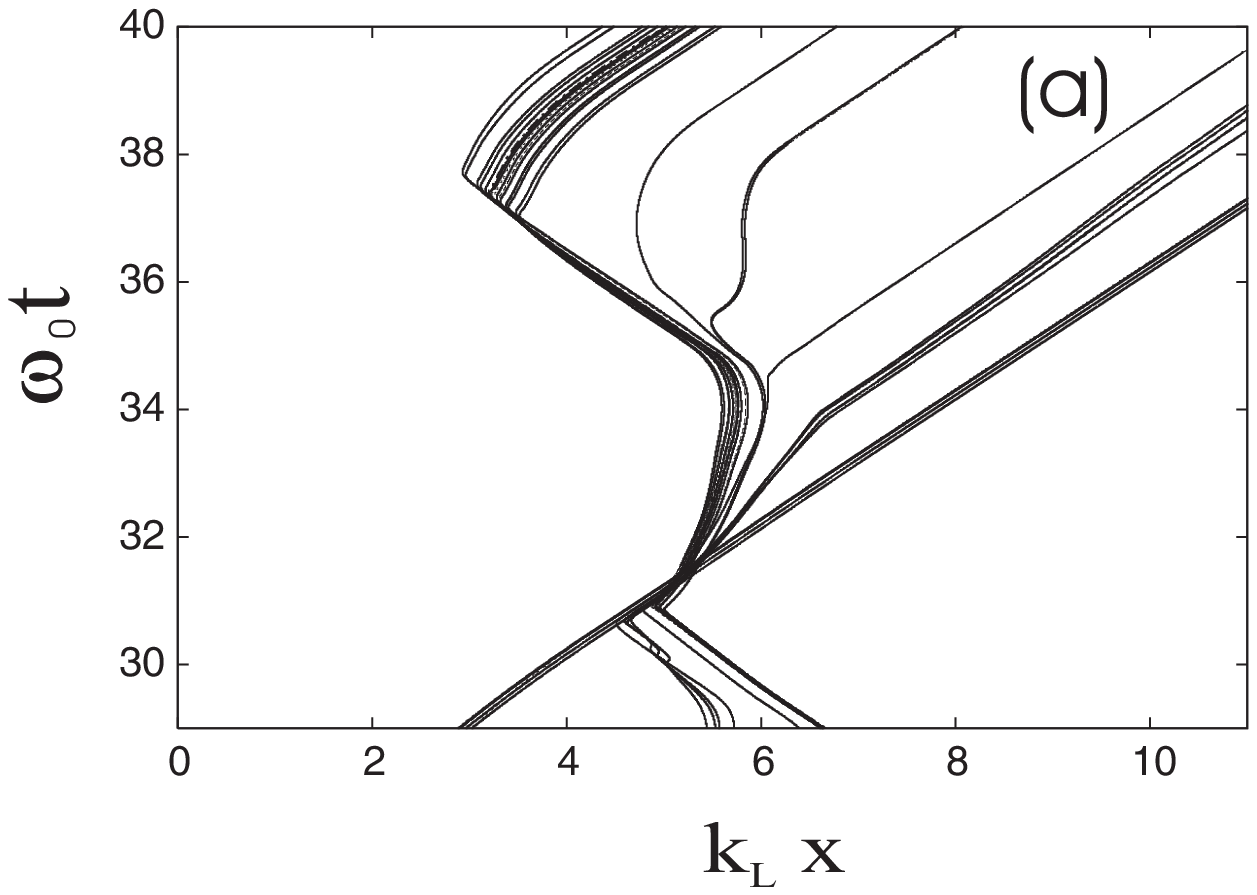} \hspace*{1.0cm}   \vspace*{-0.3cm} 
\epsfxsize=6.0cm \epsfysize=5.0cm \epsffile{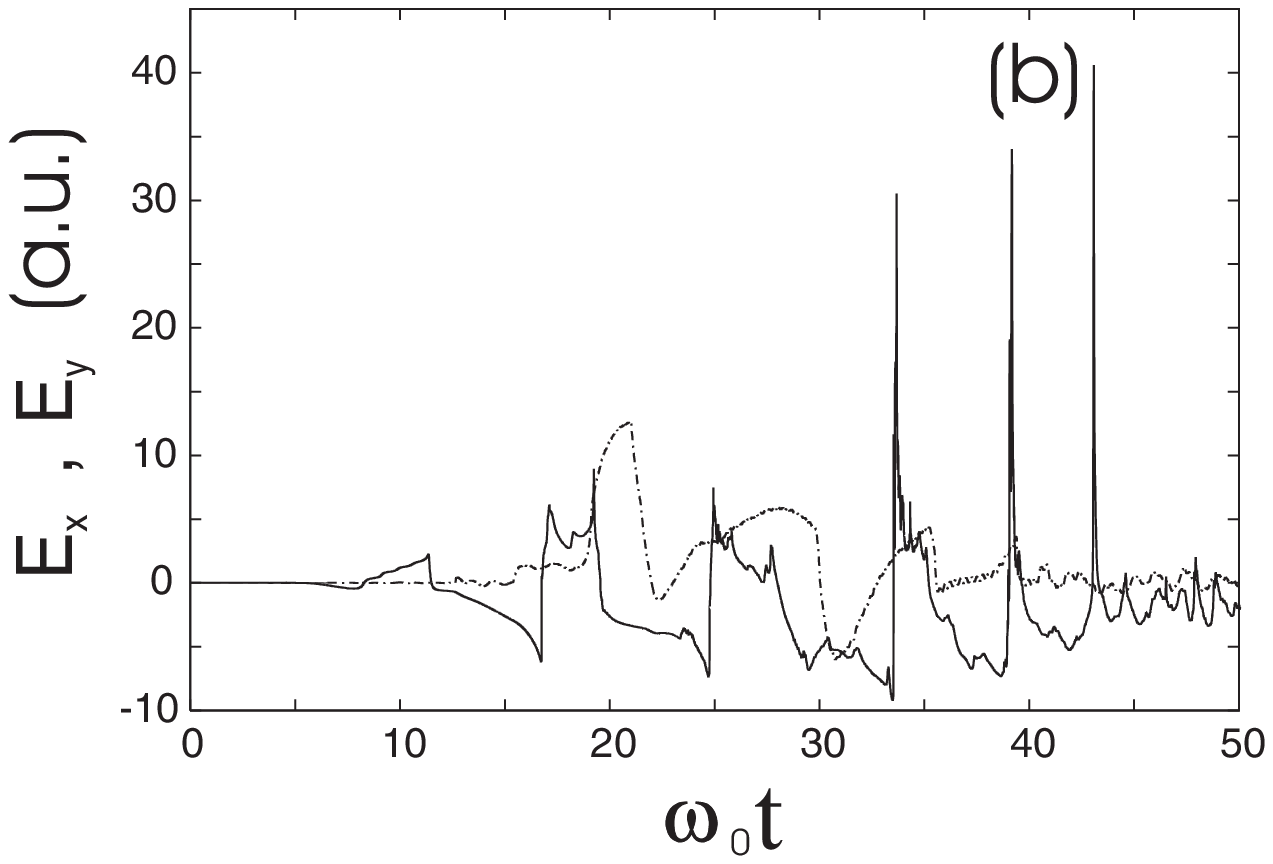} \vspace*{-0.1cm} 
\parbox{13.0cm}{\vspace*{0.5cm}Figure 2: Correlation of the electrostatic field generated in the plasma 
($E_x$,  dotted line) with: (a) Brunel electron  trajectories and beam crossings, and (b) the reflected attosecond 
electric spikes ($E_y$, solid line).}
\end{figure}

\

\noindent duced  by the scattering of the electron beams that propagate inside the plasma through regions of
 highly intense localized electrostatic fields. These localized electrostatic structures cause the beam electrons to be
 scattered and to emit  distinctive radiation. The WB model considers a specific shape and size of the
 soliton in which the characteristic turbulent density oscillation has a dipole structure within a localized region
  
\begin{equation}\label{e1}
\rho = \rho_0\,  \frac{ \hat{\bm p} \cdot {\bm r} }{D}\, \exp \left( -\frac{r^{2}}{D^{2}} \right) 
\exp \left( -i \, \omega_p\, t \right)\,\, ,
\end{equation}

\
   
 \noindent here $\rho_0$ denotes the amplitude of the density oscillation localized by a Gaussian envelope of scale
 length $D$, oscillating at the plasma frequency $\omega_p$, with a dipole moment oriented in the direction of
 $\hat{{\bm p}}\,$. It can be showed that the electric field strength near the centre of the soliton is
   
\begin{equation}\label{e2}
{\bm E_{\,s}} = -  \frac{2}{3}\, D\, \rho_0\, \exp \left( - i\, \omega_p\, t \right) \,\, {\hat{\bm p}}\,\, .
\end{equation}

\
     
 The acceleration of a charged particle through a soliton field generates electromagnetic wave packets containing a
 broad  spectrum of frequencies. The electrons in an uniform electron beam would radiate with no  phase coherence,
 resulting in  weak scattered emission. In contrast,  an electron beam with nonuniform density can maintain a degree 
of phase coherence over a range of wavelengths comparable with the scale length of density fluctuations. Therefore,
 the superposition of coherent radiation from plasma electrons can  result  in strong emission if the beam is modulated 
or bunched. Furthermore, when the beam is ultrarelativistic the radiation could be greatly enhanced by relativistic
 beaming along the  direction of propagation of the laser wave. Beam radiation would be expected to dominate plasma
emission when the beam density is high and the electrons highly relativistic. 

Thus electron bunching results in a state of strong turbulence within the scale of the localized electrostatic field and a
 broad spectrum of density fluctuations develop. Beam  density fluctuations produce regions of high $E_{s}$ fields
 which would represent centres of electromagnetic radiation producing a plasmon condensate spectrum with a 
power-law index of $p=5/3$. It can be demonstrated that the spectrum of beam density correlations  leads to a
structure function that gives the total energy radiated by the soliton, in the form
 
 \begin{equation}\label{e3}
\displaystyle  \frac{d E (\omega )}{d\, \omega} \sim V \left( \frac{\omega - \omega_p}{v_0} \right) \,\, ,
 \end{equation}
 
 \
 
 \noindent where $v_0$ is the electron beam velocity and $V$ the beam density correlation, given by [9]
  
 \begin{equation}\label{e4}
\displaystyle  V(k) \sim \frac{\Gamma (\nu + 1/2)}{\sqrt{\pi}\, \Gamma(\nu) }\, \frac{r_0}{{\left( 1 + k^{2} r_0^{2} \right)}^{\nu + 1/2}}\,\, ,
 \end{equation}
 
 \
 
 \noindent where $\Gamma$ is the gamma function, $r_0$ the perturbation scale length and the wave number 
 $k=(\omega - \omega_p)/v_0$. The  turbulent spectrum  of density fluctuations gives coherence to the emission 
 spectra when the electron beam is perturbed by  scattering density centres.  

By integrating (4) we found that the power-law index of $p=5/3$ that governs the spectrum decay is retrieved
when the parameter $\nu=1/3$.  Figure 3(a) shows the function $V(k)$  representing the beam radiation power
characterized by a prominent line at the plasma frequency, a power spectral decay 5/3 over high harmonic orders 
and an  sudden cut-off representing an upper bound at the highest frequencies emitted by the  electrons that are
scattered by the solitonic structure. 

\
 
 \renewcommand{\thefigure}{3}
\begin{figure}[h] 
\centering \leavevmode
{\epsfxsize=6.0cm \epsfysize=5.0cm \epsffile{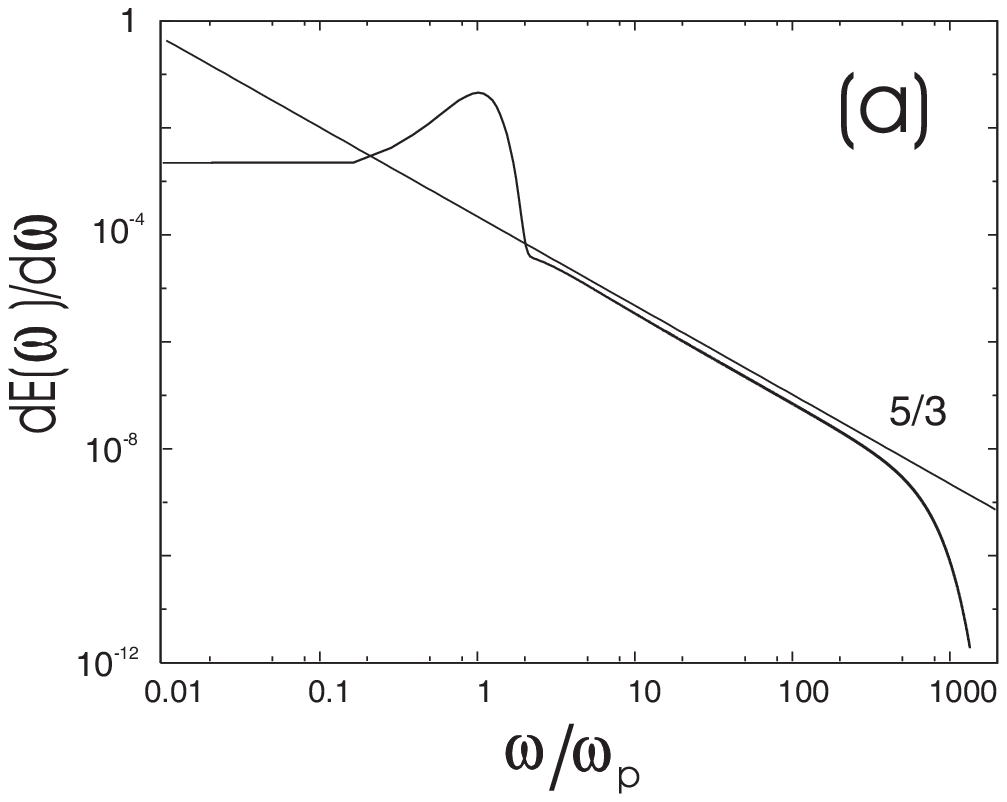} \hspace*{1.0cm} }
{\epsfxsize=6.0cm \epsfysize=5.0cm \epsffile{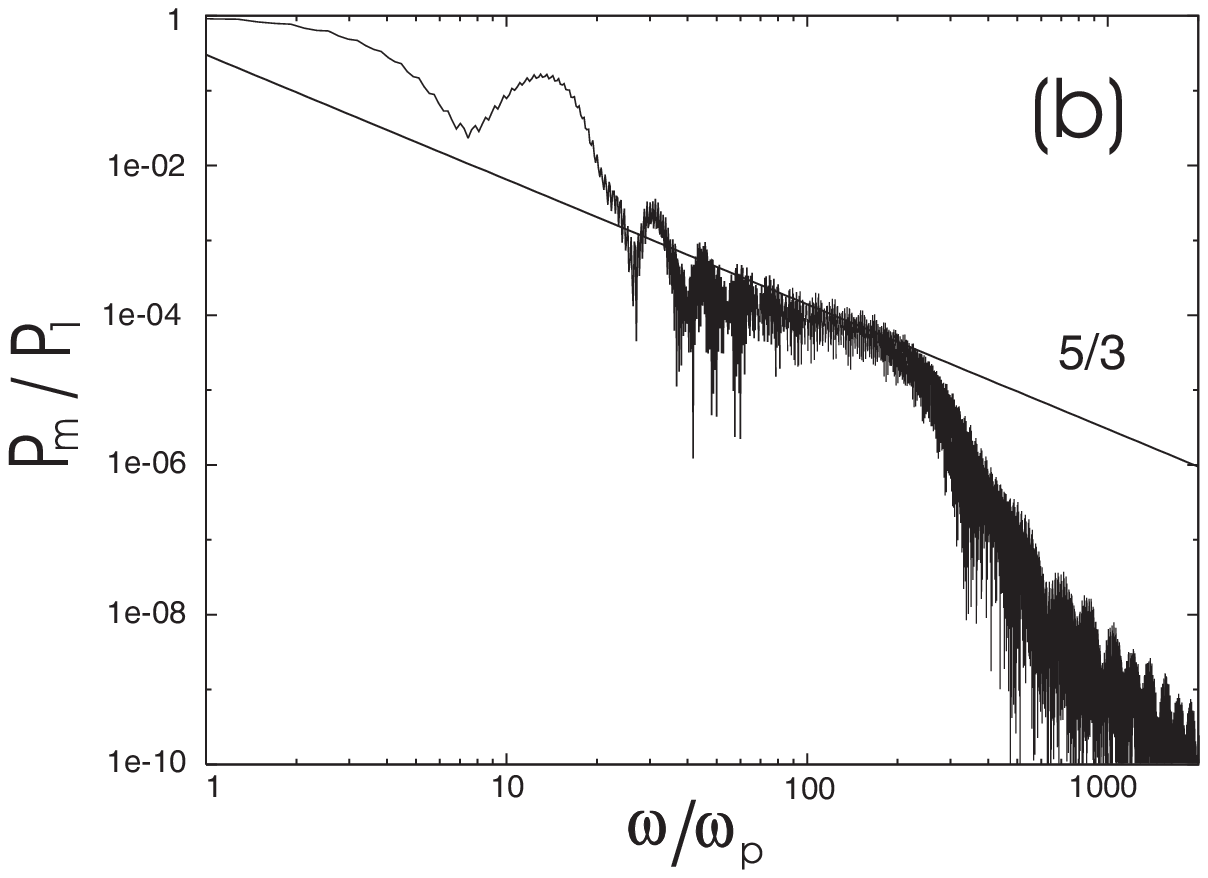} \vspace*{-0.2cm} }
\parbox{13.0cm}{\vspace*{0.5cm}Figure 3: (a) Beam radiation power spectrum from density correlations 
$V(k)$ as given by Eq.\ (4), and (b) single particle radiation from a laser-accelerated electron perturbed by a 
soliton field (Eq.\ (2)), for $a_0=40,  n_e/n_c=167$, both having power-law index decays $p \sim 5/3$. The
electron dynamics was integrated from Eq.\ (5).}
\end{figure}

\
 
Thus, the XUV spectral decay is described by a power law $  P(\omega) \sim\omega^{-5/3}$, below an abrupt cut-off at
 a critical frequency $\omega \lesssim \omega_{c} = 2\,  \gamma_{b}^{2}\, c/D$, where $\gamma_{b}$ is the
 relativistic factor for the electron beam [8]. 

By integrating the Lorentz equation we determine the motion of an electron plasma relativistically accelerated by a 
short highly intense field of a laser pulse. The dynamics was obtained from the relativistic equations for the energy 
and momentum [10]

\

\begin{equation}\label{e5}
\begin{array}{rl}
{\displaystyle  m\,c\, \dot{\gamma} = } & { \displaystyle \bm{ \beta}\, \cdot \, \bm{\dot{p}} \,\, ,} \\*[0.6cm]
{\displaystyle \bm{\dot{\beta}}\, =} & { \displaystyle \frac{e}{m\,c \,\gamma}\, \left[ \, \bm{E} - \bm{\beta}\, 
\left( {\bm E} \cdot \bm{\beta}\,\, \right) + \bm{\beta}\, \times \bm{B}\,\, \right]  + {\bm{E}}_{\, s}\,\,\, ,} 
\end{array}
\end{equation}

\vspace*{0.5cm}

 \noindent where $\gamma = {\left( 1 - \beta^{2} \right)}^{-1/2}$ is the relativistic factor, with 
${\bm \beta}={\bm v}\,/c$ 
 the normalized  electron velocity and $c$ the speed of light in vacuum. The electric and magnetic fields of the 
monochromatic plane-polarized wave are given by $\displaystyle \bm{E}=-\partial \bm{A}\,/ \partial t$  and $\bm{B}= 
\bm{\nabla}\, \times \bm{A}\,$, respectively, where $\displaystyle \bm{A}\left( \eta \right) = A_p\, P(\eta) \left[ 
\bm{\hat{x}}\,\, \delta\, \cos \eta + \bm{\hat{y}}\,\,{(1-\delta^{2})}^{1/2} \sin \eta \right]$ is the
electromagnetic potential and $\eta=\omega_L t - k_L x$, the phase of the wave with propagation along the $x$
direction. $A_p, P(\eta)$ and $\delta$ are the amplitude, the shape pulse and the polarization parameter, respectively.
The relativistic dynamics of an electron is characterized by strong nonlinearities that transform the periodic motion of a
driven harmonic oscillator into irregular motion, giving rise to very high order harmonics in the radiation spectra. A
typical spectrum is shown in Fig.\  3(b) for the combination $(a_0, n_e/n_c)=(40, 167)$. The spectrum is characterized
not only by a prominent emission at the plasma line but also by a spectral 5/3-decay.

\newpage

\vspace*{-0.6cm}

 \renewcommand{\thefigure}{4}
\begin{figure}[h] 
\centering \leavevmode
{\epsfxsize=12.0cm \epsfysize=7.5cm \epsffile{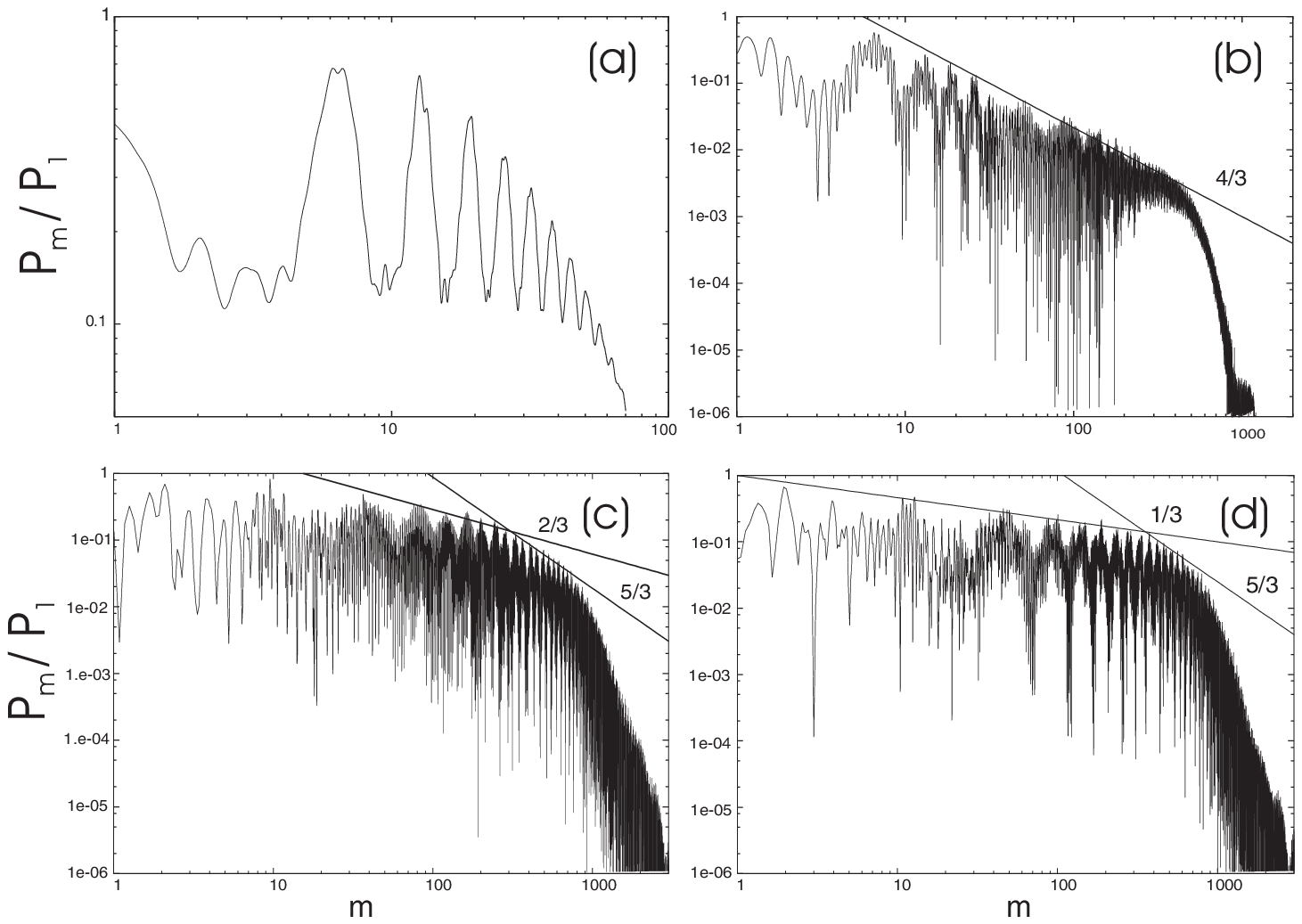} \hspace*{1.0cm} }
\parbox{12.0cm}{\vspace*{0.5cm}Figure 4: Plasma harmonic excitation for (a) $a_0=10, n_e/n_c=40,
 t_p/\tau_L=15$, (b)  $a_0=30, n_e/n_c=40,  t_p/\tau_L=10$, (c) $a_0=40, n_e/n_c=80, t_p/\tau_L$=10, and 
(d) $a_0=80,  n_e/n_c=120, t_p/\tau_L=10$.}
\end{figure}

Figures 4 (a-d) show spectra obtained from different combinations of the parameters $(a_0, n_e/n_c)$, and where 
emission at the plasma line, and its harmonics, are clearly observed together with a harmonic power decay index in the
range 2/3-5/3 at high orders. Some modulation around the plasma emission is also discerned 
in these simulations with  a cut-off at a high harmonic order as predicted by the parameter $\omega_c$. 

\

\noindent {\large \bf 4.\ Conclusions}

\

\noindent The formulated single particle radiation model is not inconsistent with our previous results obtained from
PIC simulations, with the observed highly localized electrostatic fields generated inside the plasma during the
 interaction, that correlate with the electron trajectory crossings produced when strongly accelerated beam-plasma
electrons -the so-called Brunel electrons- are re-injected into the target by the laser electric field after performing
vacuum  excursions outside the plasma. The radiation from perturbed plasma electrons by a soliton field representing
the electrostatic fields in the plasma is reminiscent of  the 5/3 UR power law decay observed in PIC emission spectra.

\

\noindent {\large \bf Acknowledgments}

\vspace*{0.2cm}

\noindent One of us (R.O.R.) acknowledges support from CONACyT under Contract No.\ 167185.

\vspace*{0.4cm}

\noindent {\large \bf References}

\vspace*{0.4cm}

\noindent [1] Teubner U and Gibbon P 2009 {\it  Rev.\ Mod.\ Phys.\ }{\bf 81} 445

\noindent [2] Norreys {PA \it et al.\ }1996 {\it Phys.\ Rev.\ Lett.\ }{\bf 76} 1832, Gibbon P 1996 {\it  Phys.\ Rev.\ 
                      Lett.\ }{\bf 76} 50

\noindent [3] Dromey B {\it et al.\ }2006 {\it  Nat.\ Phys.\ }{\bf 2} 456

\noindent [4] Baeva T, Gordienko S and Pukhov A 2006 {\it  Phys.\ Rev.\  E }{\bf 74} 046404

\noindent [5] Boyd TJM and Ondarza-Rovira R 2008 {\it Phys.\ Rev.\  Lett.\ }{\bf 101} 125004;
                      2008 {\it Laser-Driven Relativistic \hspace*{0.5cm} Plasmas, AIP Conf.\ Proc.\ }{\bf 1024} 233 

\noindent [6] Boyd TJM and Ondarza-Rovira R 2000 {\it  Phys.\ Rev.\ Lett.\ }{\bf 85} 1440, {\it idem} 2010 {\it  Phys.\
                      Lett.\ A} {\bf 374} 1517, \hspace*{0.5cm} {\it idem} 2012 {\it Phys.\ Rev.\ E} {\bf 86} 026407

\noindent [7] Weatherall JC and Hobbs WE 1986 {\it Phys.\ Fluids} {\bf 29} 2292

\noindent [8] Weatherall JC and Benford G 1991 {\it Astrophys.\  J.\ }{\bf 378} 543

\noindent [9] V.I.\ Tatarski 1967 {\it Wave Propagation in a Turbulent Media} (New York: Dover)

\noindent [10] T.J.M.\ Boyd and J.J.\ Sanderson 2003 {\it The Physics of Plasmas} (UK: Cambridge University Press)
\end{document}